# Photonic simulation of topological excitations in metamaterials

Wei Tan[†], Yong Sun[†], Hong Chen[††]
*Key Laboratory of Advanced Micro-structure Materials, MOE, Department of Physics, Tongji University, Shanghai 200092, China*
Shun-Qing Shen[††]
*Department of Physics, The University of Hong Kong, Pokfulam Road, Hong Kong, China*

[†]These authors contributed equally to this work.
[††] Correspondence and requests for materials should be addressed to H. C. or S. Q. S..

Abstract: Condensed matter systems with topological order and metamaterials with left-handed chirality have attracted recently extensive interests in the fields of physics and optics. So far the two fields are independent, and there is no work to address their connection. Here we propose to establish the relation between the topological order in condensed matter systems and the chirality in metamaterials, by mapping explicitly Maxwell's equations to the Dirac equation in one dimension. We report an experimental implement of the band inversion in the Dirac equation, which accompanies change of chirality of electromagnetic wave in metamaterials, and the first microwave measurement of topological excitations and topological phases in one dimension. Our finding provides a proof-of-principle example that electromagnetic wave in the metamaterials can be used to simulate the topological order in condensed matter systems and quantum phenomena in relativistic quantum mechanics in a controlled laboratory environment.

The Dirac equation provides a description of relativistic quantum mechanics for an elementary spin-1/2 particle [1],[2], which predates the discovery of positron, an anti-particle of electron in high energy physics [3], and also has extensive applications in condensed matter systems such as graphene [4],[5] and topological insulators [6]–[8]. Recent years it is realized that it is a key to understand topological phases from one- to three-dimensional systems and from insulators to superconductors or superfluids. On the other hand, Maxwell's equations form the foundations of classical electrodynamics and modern optics. Modern techniques and material sciences make it possible to precisely control photonic transport in artificial optical materials or metamaterials, which attracts great interests in the field of optics. Photonic technology available now is reaching at the stage to illustrate unusual ways for photon manipulation such as negative refraction [9],[10], electromagnetic cloaking [11][12], structure-induced coherence [13–16], and mimicking photonic black holes [17]. Here we propose to link the metamaterials to topological phenomena in condensed matter systems and the relativistic quantum mechanics. We find that the one-dimensional (1D) Maxwell's equations can be written in the compact form which has the identical mathematical structure of the Dirac equation. We perform a proof-of-principle photonic simulation of the Dirac equation in metamaterials by means of the full wave numerical simulation and microwave experiment of transmission line. For the first time we successfully implement the band inversion of the Dirac equation. It is noted that the band inversion accompanies changing chirality of electromagnetic wave in metamaterials from the right-handed to left-handed triad, which determines a matter-antimatter correspondence in the relativistic quantum mechanics. Furthermore we utilize designing metamaterials to observe experimentally the topological phases and excitations in one dimension. This paves the way to investigate the topological phenomena in condensed matter systems and the Dirac-like particles in high-energy physics in a photonic simulator with controlled experimental parameters [18]. Meanwhile we can also make use of the solutions of the Dirac equation to understand exotic phenomena observed in metamaterials.

**Photonic analog to the Dirac equation** – A 1D plane electromagnetic wave of the frequency $\omega$ in an optical media can be described by Maxwell's equations

$$-\partial_x E_z = i\omega\mu_0\mu_r(x)H_y, \quad (1)$$

$$\partial_x H_y = -i\omega\varepsilon_0\varepsilon_r(x)E_z. \quad (2)$$

Here, $E_z$ is the electric field and $H_y$ is the magnetic field. $\varepsilon_0$ and $\varepsilon_r$ ($\mu_0$ and $\mu_r$) are the vacuum and dimensionless relative permittivity (permeability) of the media, respectively, and can be functions of position in designed artificial optical materials. By introducing the spinor $\varphi = \begin{pmatrix} \sqrt{\varepsilon_0}E_z \\ \sqrt{\mu_0}H_y \end{pmatrix}$, Eqs. (1) and (2) can be written as,

$$[-i\sigma_x\partial_x + m(x)\sigma_z + V(x)]\varphi = E\varphi. \quad (3)$$

Here $m(x) = \frac{\omega}{2c}(\varepsilon_r - \mu_r)$ and $V(x)$ are the effective mass and potential, respectively. $E$ is the energy eigenvalue and $V(x) - E = \frac{\omega}{2c}(\varepsilon_r + \mu_r)$. Here $c$ is the speed of light in vacuum and $\sigma_{x,y,z}$ are the Pauli matrices. Eq. (3) is equivalent to the stationary Dirac equation in a potential $V(x)$ by taking the Planck constant $\hbar$ and $c$ as units. In this way we have established a one-to-one mapping between Maxwell's equations and the Dirac equation in 1D, which provides a platform to study relevant problems of the Dirac equation in metamaterials with engineered permittivity and permeability.

It is noted that the effective mass in Eq. (3) is given by the difference between the permittivity and permeability, which can be tailored by artificial structures in metamaterials. The signs of permittivity and permeability determine the chirality of electromagnetic wave in optical media: the electric field, magnetic field and the wave vector obey the right-handed or left-handed triad. For ordinary (nonmagnetic) optical materials $\varepsilon_r > \mu_r = 1$, the electromagnetic wave obeys the right-handed rule with a positive effective mass $m > 0$ while the double-negative or left-handed metamaterials with $\varepsilon_r < \mu_r < 0$ [9] have a negative mass $m < 0$ and obey the left-handed rule. On the other hand, it is known that the sign of the effective mass is closely related to the topological order of a medium. Therefore, the present study is the first one, to the best of our knowledge, to illustrate the close relation between topological order in condensed matters and chirality feature in metamaterials. We believe that this relation will open a new route to mimic topological phases and excitations of Dirac equation in designed metamaterials, and to understand some exotic phenomena in metamaterials from a point of view of the Dirac equation.

**Simulation of the band inversion in the Dirac equation** – The Dirac equation demands the existence of anti-particle, the particles with negative energy and negative mass. The solutions of positive and negative energy automatically satisfy the Einstein mass-energy relation as a consequence of special theory of relativity. To interpret the solutions, Dirac proposed that the negative energy solution is for a positron with negative mass, an anti-particle of electron [1],[2]. According to the Pauli exclusion principle, an electron cannot occupy the state of negative energy as all the states with negative energy are supposed to be fully filled. There exists an energy gap $2m_e c^2$ ($m_e$ is the rest mass of electrons) between the positive energy band for an electron and the negative energy band for positron. As the rest mass of electron is very huge, $m_e c^2 = 0.53$ MeV, a positron can only be observed in high energy physics. However, the mapping between Maxwell's equations and the Dirac equation in 1D offer an alternative approach to realize the band inversion, *i.e.*, the sign change of effective mass in the Dirac equation, in metamaterials because either the permittivity $\varepsilon_r$ or permeability $\mu_r$ in metamaterials can be manipulated in a controllable way. It was reported that a periodic array of two kinds of single-negative metamaterials, in which either permittivity or permeability is negative, with sub-wavelength unit cell shows a novel photonic band gap at the center of wave number $k=0$ [19]. The lower band presents a negative-refractive-index, while the band above has a positive index, which provides a controllable way to manipulate the sign of the effective mass $m$ in Eq. (3).

Specifically, to bridge the photonic gap and the Dirac gap, we calculate the dispersion relation from the perspective of the Dirac equation, as shown in Eq. (3). By treating metamaterials with sub-wavelength unit cell as an effective media with complex electric and magnetic responses, we can obtain the dispersion relation,

$$k^2 = (V-E)^2 - m^2 = \frac{\omega^2}{c^2}\varepsilon_r \cdot \mu_r. \qquad (4)$$

Here $k$ is real if $\varepsilon_r \mu_r > 0$, corresponding to either the positive-index band with right-handed chirality or negative-index band with left-handed chirality, respectively. In contrast, $k$ has a purely imaginary value if $\varepsilon_r \mu_r < 0$, which also indicates the existence of a band gap because the electromagnetic wave will decay exponentially in the media. It follows from Eq. (3) that the gap can be characterized as a positive mass $m > 0$ ($\varepsilon_r > \mu_r > 0$) and a negative one $m < 0$ ($\varepsilon_r < \mu_r < 0$), respectively.

Microwave experiments based on transmission-line (TL) metamaterials are performed to realize the band inversion from a positive to negative mass. The TLs are all fabricated on copper-clad 1.57-mm thick Rogers RT5880 substrates. A network analyzer (Agilent PNA N5222A) was used to characterize our samples in frequency domain. Transmission and reflection properties were obtained directly, and the density of states is calculated from the measured group delay. Then we carried out microwave experiments in time domain to investigate the field distribution. At the particular frequency, a monochromatic wave generated from Agilent E8267D is input to the sample. After that, voltage signals at the different positions along the sample are picked and recorded, using the high-impedance active probe (Tektronix P7260) and the oscilloscope (Tektronix TDS7704B). In each unit cell, only one position which is near the shunt inductor is probed, and thus the *LC* resonances within a unit are not detected. Numerical simulations were obtained using a commercial software package (CST Microwave Studio). Full methods, specific parameters of samples, and experiment measurement are available in the Supplementary Information.

In our designed samples, the effective permittivity and permeability of the TL metamaterials are given by [20],[21]

$$\bar{\varepsilon}_r = \frac{1}{p\varepsilon_s}\left(C_0 - \frac{1}{\omega^2 Ld}\right) + i\frac{\gamma_e}{\omega}, \quad \bar{\mu}_r = \frac{p}{\mu_s}\left(L_0 - \frac{1}{\omega^2 Cd}\right) + i\frac{\gamma_m}{\omega}, \tag{5}$$

where $\varepsilon_s$ and $\mu_s$ are the permittivity and permeability of environment media, $C_0$ and $L_0$ are the per-unit-length capacitance and inductance of the TL segment, respectively, $C$ and $L$ are the series capacitance and the shunt inductance of the loading elements, $\gamma_e$ and $\gamma_m$ denote the losses, $d$ is the length of a unit cell, and $p$ is the geometric factor. We would like to emphasize that these parameters can be tailored experimentally. For example, we can tune the frequency $\omega$ or the length of a unit cell $d$ to change the values of the permittivity and permeability continuously. From the frequency-dependent $\varepsilon_r$ and $\mu_r$, the two band edges $\omega_1$ and $\omega_2$ are determined by setting $\varepsilon_r = 0$ or $\mu_r = 0$, respectively. The frequency difference between $\omega_1$ and $\omega_2$ defines the energy gap as shown in Fig. 1a. The gap closing at $\omega_1 = \omega_2$ or $m = 0$ gives rise to the Dirac point, where some interesting behaviors have been reported, such as Zitterbewegung [22], wave bending and cloaking effect [23].

The band inversion is illustrated in the simulated and measured density of states (DOS) in Fig. 1. Agilent PNA Network Analyzer N5222A was employed to measure the reflection, transmission, and group delay of the samples. The DOS can be deduced from the measured group delay $\tau_g$, $g(\omega) = \frac{1}{\pi}\frac{dk}{d\omega} = \frac{\tau_g}{\pi D}$ (where $D$ is the total length of the sample), which reveals the band structure of the system. With an increase of the width of the TL $w$, $C_0$ increases, but $L_0$ decreases. Consequently, one band edge moves to a higher frequency, while the other moves to a lower one. Therefore in a well-designed structure, the band inversion is expected to occur by adjusting $w$. We designed several samples and calculated their width-dependent dispersions, and observed the band inversion as shown in Fig. 1a. It is noted that the band inversion accompanies an exchange of the band edges of $\varepsilon_r = 0$ or $\mu_r = 0$, and the effective mass also changes its sign. This feature is closely related to the chirality of electromagnetic wave in the designed sample as shown in Fig. 1b. Before the band inversion, the edge of $\mu_r = 0$ locates at higher frequency and connects to the pass band with right-handed chirality for $\varepsilon_r > 0, \mu_r > 0$ while the edge of $\varepsilon_r = 0$ locates at lower frequency and connects to the bass band with left-handed chirality for $\varepsilon_r < 0, \mu_r < 0$, as shown in the Fig. 1b (i). After the band inversion from $m > 0$ to $m < 0$, the two band edges are exchanged with each other as shown in Fig. 1b (iii): the edge of $\mu_r = 0$, for example, locates at lower frequency and connects to the pass band with left-handed chirality for $\varepsilon_r < 0, \mu_r < 0$. It is known that, the sign of the effective mass in Dirac equation has been used to describe different topological order in condensed matter systems [24],[25]. Therefore, the above picture shows us a clear connection between topological order in condensed matter systems and chirality in metamaterials.

To illustrate the band inversion experimentally we designed and fabricated a series of samples and measured the DOS for microwaves. Meanwhile we also simulated the DOS numerically with a fitted loss from experimental data. The numerical and experimental results are presented in Figs. 1c and 1d, respectively. It is clearly shown that the band gap gradually closes up as $w$ decreases to a critical point of $w_0 = 4.5$ mm. With a further decrease of $w$, a gap re-opens again. The measured data are in a good agreement with the numerical simulation. In this way we have successfully demonstrated the band inversion in the metamaterials experimentally. This provides an explicit and solid foundation for photonic simulation of the Dirac equation in metamaterials.

**Soliton solution for a domain wall** – The Jackiw-Rebbi solution describes the bound state of a particle to the interface or domain wall between two media with positive and negative masses [26]. For simplicity consider two 1D media with positive mass $m_1 > 0$ and negative mass $-m_2 < 0$ forming a domain wall at $x = 0$ with a potential $V(x) = 0$. It is found that there exists an analytical solution of zero energy ($E = 0$),

$$\varphi(x) = \sqrt{\frac{m_1 m_2}{m_1 + m_2}} \begin{pmatrix} 1 \\ -i \end{pmatrix} \exp[-|m(x)x|], \qquad (6)$$

which decays exponentially in $|x|$. It is a solution of 1D topological excitation or soliton, and has potential applications in condensed matter physics. For example, the charge carriers in 1D organic conductors are attributed to the solitons and anti-soliton [27],[28]. However, it is still an experimental challenge to observe a single soliton in a 1D condensed-matter system due to small lattice spacing [29],[30].

Here we demonstrate that the Jackiw-Rebbi solution can be realized in the metamaterials by constructing a domain wall with controllable parameters. To this end, we fabricated a sample consisting of two TL metamaterials, $w_1 = 2.5$ mm and $w_2 = 8.5$ mm. Both metamaterials have finite gaps, but with opposite masses as illustrated in Figure 1c and 1d. However, when these two TL metamaterials are connected, it is found that an additional narrow peak appears at $\omega_0 = 11.05$ GHz within the gap region in the DOS as shown in Fig. 2a. Figure 2b shows the full-wave simulation of field spatial distribution in the sample at $\omega_0$. The measured field distribution is presented in Fig. 2c. The results indicate that the incident field increases to reach a maximal, and then decays exponentially. Usually the incident field should decrease exponentially in a band gap for a material with loss. The measured peak of the incident field is attributed to the non-linear or topological excitation described by Jackiw-Rebbi solution near the interface. Thus this measurement provides the first direct observation of the Jackiw-Rebbi solution or the profile of soliton in a photonic simulator made of metamaterials.

**Simulation of a 1D lattice topological phase** – The metamaterials with different effective masses provide building blocks to construct various artificial optical materials to simulate solid state systems. In condensed matter, the simplest "two-band" model is the Su-Schrieffer-Heeger model for polyacetylene [27]. Consider a 1D dimerized lattice with bipartite lattice sites A and B. Each unit cell consists of two sites A and B. The hopping amplitude in between two sites in a unit cell is $t + \delta t$ and that between two unit cells is $t - \delta t$. When $\delta t = 0$, the energy dispersion $\pm 2t \left| \cos \frac{k}{2} \right|$, which exhibits no energy gap at $k = \pi$. However at half filling, due to the Peierls' instability, the dimerization occurs and $\delta t \neq 0$: an energy gap opens, and is equal to $4\delta t$. According to the sign of $\delta t$, the gap can be either positive or negative. The inverted band structure is closely related to topological insulators, such as $Bi_2Te_3$ and $Bi_2Se_3$ [31],[32]. The topological property of this 1D model also reveals form the Berry phase of the lower energy band, which is given by $\pi[(\text{sgn}(t + \delta t) - \text{sgn}(\delta t)]/2$ modulus by $2\pi$. The difference of the Berry phases for $\delta t > 0$ and $\delta t < 0$ is π, which indicates that the two phases are topologically distinguished: one is topologically trivial and the other one is topological non-trivial. The topologically non-trivial phase is characterized by the presence of end state of zero energy in an open boundary condition [33][34]. Though it is believed that the end states should exist in 1D polymer, it is a great challenge for experimentalists to measure them experimentally.

To mimic topological phases in a lattice structure, we design a periodic stack of two TL blocks with $m_A > 0$ ($w = 7$ mm) and $m_B < 0$ ($w = 4$ mm) as shown in Fig. 3a. This periodic structure can be used to simulate the Su-Schrieffer-Heeger model. Different $m_A > 0$ and $m_B < 0$ corresponds to the positive and negative $\delta t$. In the periodic boundary condition, the theoretical calculation shows that dispersion relation have the similar band structure of the Su-Schrieffer-Heeger model, which exhibits the presence of an energy gap $\delta\omega = 2.42$ GHz between two band edges at $\omega_1 = 12.77$ GHz and $\omega_2 = 15.19$ GHz.

In our experiment, we fabricated two structures by removing one block of $m_A > 0$ or $m_B < 0$ from the loop as shown in the inset of Fig. 3(b). The measured DOS for microwave show that the first case in Fig. 3(c) exhibits a clear gap $\delta\omega = 2.57$ GHz between band edges at $\omega_1 = 12.52$ GHz and $\omega_2 = 15.09$ GHz, which value is close to the calculated value of the loop. The non-zero DOS is attributed to the loss of the metamaterials, which is characterized by the parameters $\gamma_e$ and $\gamma_m$ in Eq. (5). They are fitted to be 0.24 from the measured data. The second case in Fig. 3(d) exhibits a similar band structure as in Fig. 3(c), but presents an additional peak at $\omega_0 = 13.71$ GHz between the two peaks at $\omega_1 = 12.22$ GHz and $\omega_2 = 15.64$ GHz. A more detailed analysis indicates that the two peaks at $\omega_1 = 12.22$ GHz and $\omega_2 = 15.64$ GHz are band edges as shown in Fig, 3(b), in which slight shifts of the position are caused by the finite size effect. The resonant peak at $\omega_0 = 13.71$ GHz corresponds to two bound states at the ends. This indicates that the topological properties of the two designed chains in Fig. 3(c) and 3(d) are topologically distinguished, although they are constructed by the same blocks of $w = 7$ mm and $w = 4$ mm. Thus our measurement demonstrates explicitly the existence of the end states in 1D topological phase.

In conclusion, we demonstrated an explicit mapping between Maxwell's equations and the Dirac equation in one dimension. This provides a platform to utilize the electromagnetic wave to mimic quantum phenomena related to the Dirac equation from high energy physics to condensed matter physics. The band inversion in the Dirac equation accompanies with a change of chirality of electromagnetic wave in metamaterials. Our numerical simulation and microwave experiments illustrated a proof-of-principle that metamaterials are idea candidates to simulate topological phenomena in solids, and the behaviors of the Dirac equation.

**Acknowledgements:** The work was supported by National Basic Research Program (973) of China Grant No.: 2011CB922001 (HC) and Research Grant Council of Hong Kong under Grant No.: N_HKU748/10 (SQS).

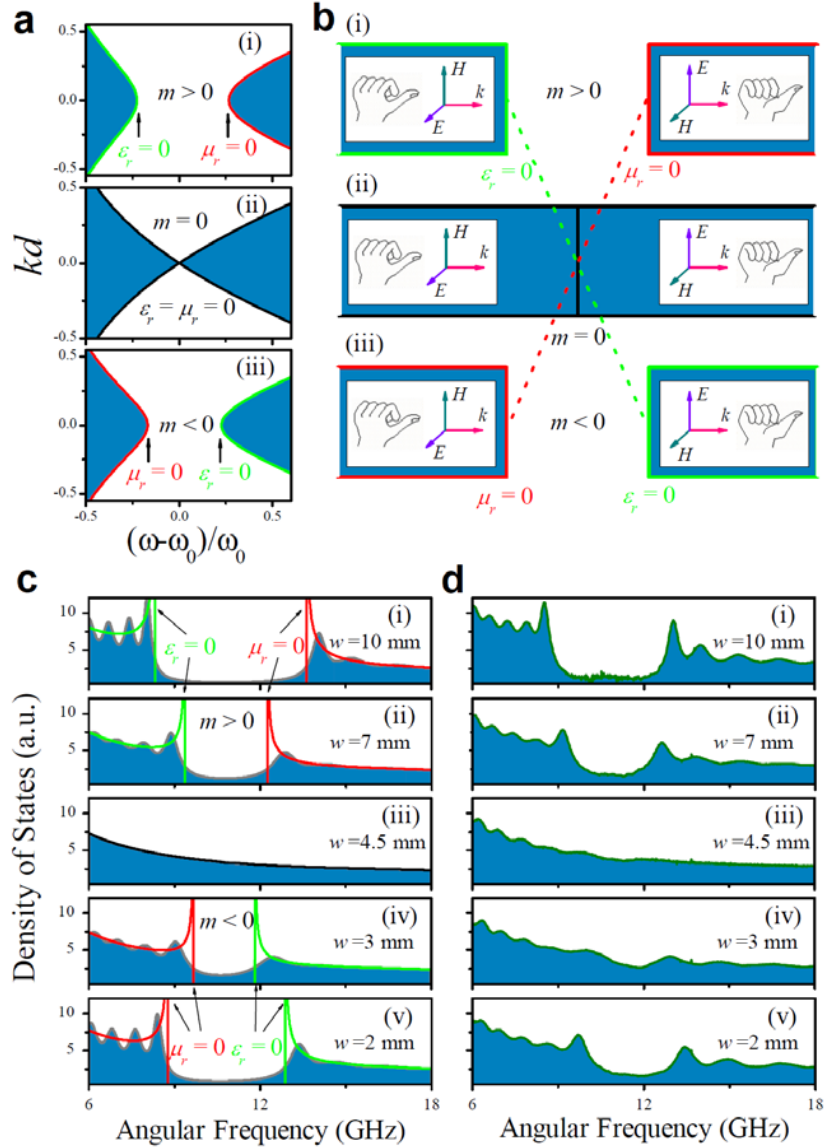

**Figure 1| Photonic simulation of band inversion in the Dirac equation. (a)** Calculated dispersion relation for the band structures with (i) $m>0$, (ii) $m=0$, and (iii) $m<0$ (iii). $\omega_0 \approx 11.34$ GHz is the frequency when the band gap closes. **(b)** The band inversion, the exchange of the band edges and change of chirality from the right-handed triad to the left-handed triad. **(c)** Calculated density of states for microwave TL samples containing 24 unit cells with fitted losses from experimental data. The solid (red) lines are the DOS spectra for ideal structures without the losses. **(d)** Measured density of

states for microwave TL samples.

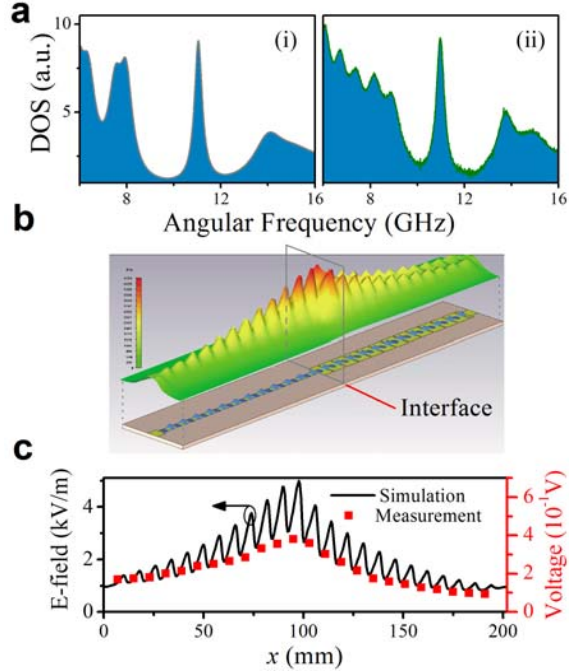

**Figure 2| Domain wall at the interface between two components with opposite masses.** (**a**) Calculated (i) and measured (ii) DOS spectra, exhibiting a bound state within the gap. (**b**) Full-wave simulation of field distribution of the bound state, presenting localized fields around the interface ($x = 96$ mm). (**c**) Measured voltage as the field distribution in the sample. The two components are designed with $w_1 = 2.5$ mm and $w_2 = 8.5$ mm, respectively. The parameters at $\omega_0 = 11.05$ GHz are given by $m_1 = -16.1$, $V_1 = 0.36 + 1.09i$, $m_2 = 22.7$, $V_2 = -0.36 + 1.09i$, and $E = -0.05$. For simplicity, we take the losses $\gamma_e = \gamma_m = 0.24$ in numerical simulation, which is fitted from experimental data. It should be noted here that in each unit cell, only one position near the shunt-inductance position is probed.

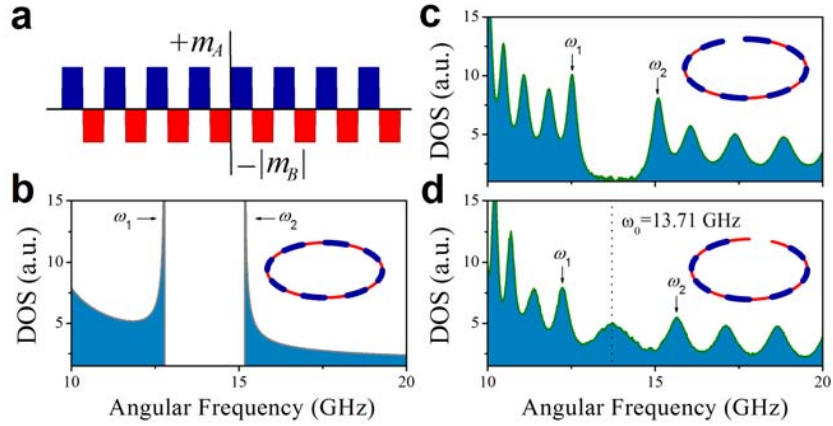

**Figure 3| Topological properties of a lattice structure.** (**a**) Scheme of the periodic structure composed of the components of $m_A > 0$ and $m_B < 0$. (**b**) DOS spectra of the ideal periodic structure. A band gap of $\delta\omega = 2.42$ GHz is present between $\omega_1 = 12.77$ GHz and $\omega_2 = 15.19$ GHz. (**c**) The measured DOS for the structure with an open boundary condition by removing one block of $m_A > 0$ component. Two band edges appear at $\omega_1 = 12.52$ GHz and $\omega_2 = 15.09$ GHz. (**d**) The measured DOS for the structure lacking of a $m_B < 0$ component. Two band edges appear at $\omega_1 = 12.22$ GHz and $\omega_2 = 15.64$ GHz, and a resonant peak for the end mode appear at $\omega_0 = 13.71$ GHz. Inset, schemes of the structures in three cases.